\documentclass{article}
\usepackage{spconf,amsmath,graphicx}
\usepackage{amssymb}
\usepackage{bm}
\usepackage{subfigure}
\usepackage{graphicx}
\usepackage{float} 
\usepackage{color}
\renewcommand{\paragraph}[1]{{\vspace{0.5mm}\noindent\normalfont\bfseries#1\quad}}
\usepackage{enumitem}
\setlist{nosep, leftmargin=14pt}

\usepackage{mwe} 

\title{Brain Cancer Survival Prediction on Treatment-na\"ive  MRI using Deep Anchor Attention Learning with Vision Transformer}
\name{Xuan Xu$^{1}$, Prateek Prasanna$^{2}$
\thanks{$^{1}$Department of Computer Science, Stony Brook University, NY }
\thanks{$^{2}$Department of Biomedical Informatics, Stony Brook University, NY}}
\address{Stony Brook University}
\begin{document}
%
\maketitle
© 2022 IEEE.  Personal use of this material is permitted.  Permission from IEEE must be obtained for all other uses, in any current or future media, including reprinting/republishing this material for advertising or promotional purposes, creating new collective works, for resale or redistribution to servers or lists, or reuse of any copyrighted component of this work in other works
\begin{abstract}
Image-based brain cancer prediction models, based on radiomics, quantify the radiologic phenotype from magnetic resonance imaging (MRI). However, these features are difficult to reproduce because of variability in acquisition and preprocessing pipelines. Despite evidence of intra-tumor phenotypic heterogeneity, the spatial diversity between different slices within an MRI scan has been relatively unexplored using such methods.  
In this work, we propose a deep anchor attention aggregation strategy with a Vision Transformer to predict survival risk for brain cancer patients. 
A Deep Anchor Attention Learning (DAAL) algorithm is proposed to assign different weights to slice-level representations with trainable distance measurements. 
We evaluated our method on N = 326 MRIs. Our results outperformed attention multiple instance learning-based techniques. DAAL highlights the importance of critical slices and corroborates the clinical intuition that inter-slice spatial diversity can reflect disease severity and is implicated in outcome.
\end{abstract}
\section{Introduction}
\label{sec:intro}

\begin{figure}
	\centering
	\includegraphics[width=0.4\textwidth]{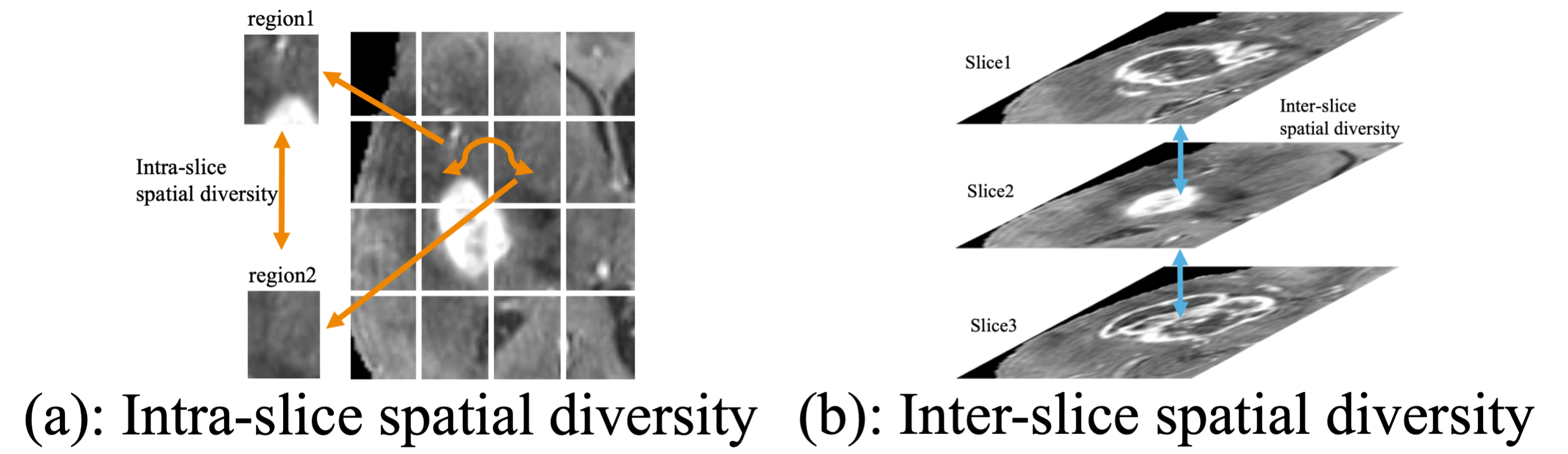}
    \label{intra}
	\caption{Illustration of intra-tumor heterogeneity.}
	\label{fig:spatial_new}
\end{figure}

\paragraph{Clinical Motivation.}
Glioblastoma (GBM), an aggressive form of brain cancer, has a poor prognosis with the median survival around 12-15 months~\cite{10.1093/brain/awm204}. This has led to an unmet clinical need for personal risk prediction approach at an early stage so that treatment regimens may be tailored according to individual risk profiles instead of resorting to a one-size-fits-all approach. MRI is a key tool in brain cancer diagnosis and treatment planning.
Previous works have leveraged  multi-parametric MRI to detect the GBM saliency ~\cite{banerjee2016novel}.
Most existing imaging models have relied on radiomic analysis~\cite{lao2017deep,choi2020multi}, which is sensitive to image pre-processing and often hard to reproduce. Though some of this heterogeneity is manifested on imaging, the spatial diversity and the interaction between different sub-regions of a tumor are not fully explored from a phenotype perspective. Survival analysis studies are limited to understanding of intra-tumoral and peri-tumoral heterogeneity as slice-wise statistics~\cite{prasanna2017radiomic} without taking into account the intra-tumor heterogeneity reflected as inter-slice and intra-slice spatial diversity on MRI (illustrated in Figure~\ref{fig:spatial_new}).


\paragraph{Technical Motivation.}
Several works have leveraged pretrained models to extract imaging features from radiological scans~\cite{pham2020comprehensive}. Pretrained models such as VGG~\cite{simonyan2014very} and ResNet~\cite{he2016deep} are usually trained on ImageNet~\cite{deng2009imagenet}. 
Pretrained AlexNet models have been used in a transfer learning setting to detect abnormalities in brain MRI~\cite{talo2019convolutional,krizhevsky2012imagenet}.
The visual composition of medical images, especially radiology scans, are fundamentally different from natural images. Pre-trained representations may not effectively capture subtle local variations prevalent in MRI scans, more so in brain tumors~\cite{raghu2019transfusion}.
Another challenge stems from the multiple views of 3D MRI scans and multiple cross-sections in a given view.
Class label is generally assigned at an MRI-level and not at a slice-level. Due to the intra-tumor heterogeneity in cancers, it is unreasonable to assign the same patient label to different slices. 


Previous works on medical imaging leveraging convolutional neural networks focus on local variations in image patterns~\cite{yao2020whole}. Vision transformer (ViT)~\cite{dosovitskiy2020image} models long range dependencies and shows how a pure transformer can perform well on image classification tasks. This closely follows the original Transformer~\cite{vaswani2017attention} and has been widely used in image classification ~\cite{han2020survey}. Compared with computer vision tasks involving natural images, MRIs are usually smaller datasets, typically including several hundred images while ViTs require large scale datasets for training.

To address these challenges, we first utilize slices from axial, coronal, and sagittal planes to extract imaging representations for each patient. Unlike existing prognostic methods that consider only axial planes~\cite{lao2017deep}, the slices from three planes enable partial retention of 3D information and a more comprehensive feature extraction with transfer learning. Secondly, we use a vision transformer~\cite{dosovitskiy2020image} to evaluate the spatial relationships between different small regions of the tumor and its periphery within one slice. Most importantly, we propose a deep anchor attention learning algorithm to exploit the inter-slice spatial diversity. To mimic the radiological assessment workflow, slices with largest tumor area are designated as \textit{anchor slices}. This strategy provides a way to aggregate the slice-level representations to patient-level representation using a latent-space comparison between anchor slices and other slices.
Our main contributions are as follows:

\begin{itemize}
    \item Intra-slice spatial diversity in MRI is captured by a self-attention mechanism. We leverage a vision transformer to quantify the spatial diversity between small regions within a given slice, as shown in Figure \ref{intra}(a). 
    \item A Deep Anchor Attention Learning (DAAL) algorithm is proposed to provide an efficient aggregation strategy for slice-level representations. 
    We adapt DSMIL~\cite{li2020dual} to highlight the critical slices in a 3D MRI. 
\end{itemize}

Our experimental results corroborate the clinical intuition that intra-tumor heterogeneity reflected on imaging as intra- and inter-slice spatial diversity (as described in Figure \ref{fig:spatial_new}) can reflect disease severity and is implicated in prognosis.

\section{Methods}
\label{sec:format}

\begin{figure*}[htb]
\centering

\includegraphics[ width=14cm]{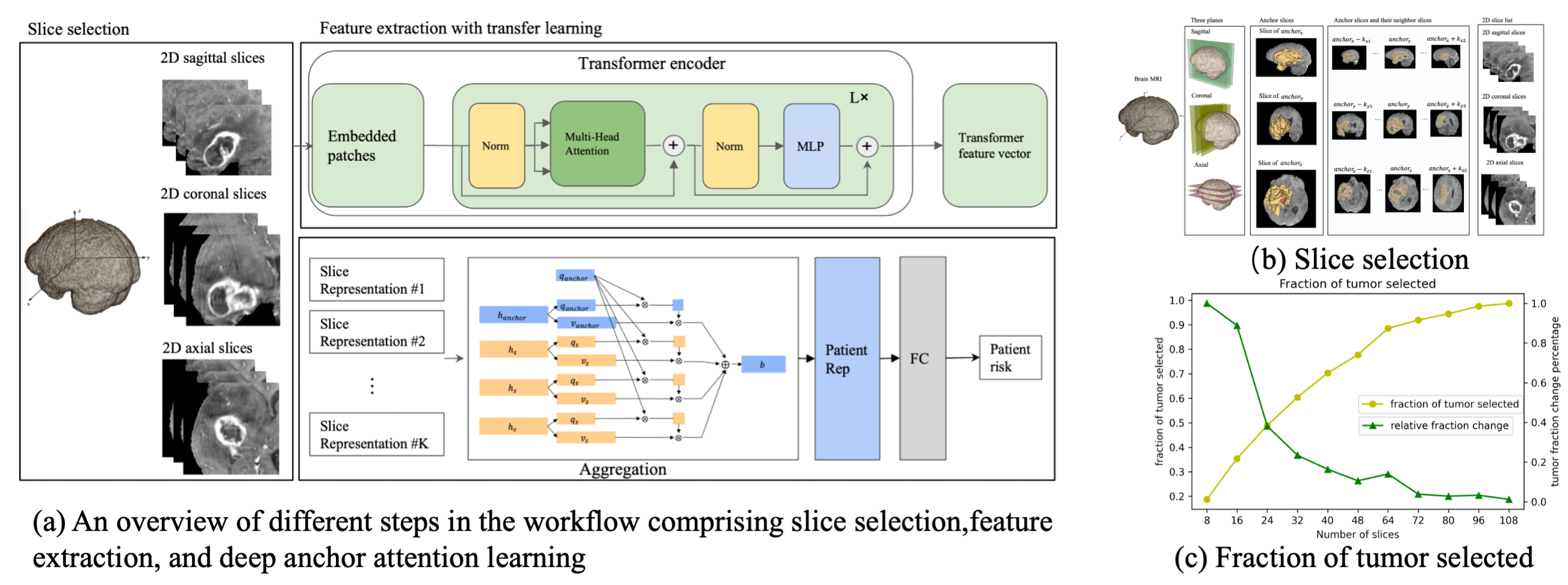} 
\caption{An overview and slice selection process.}
\label{Fig.slice}
\end{figure*}

Consider a set of N patients, ${X_i}, i=\{1,2,...,N\}$ with label $(t_i,\delta_i)$. $\delta_i$ = 1 indicates a death event and $\delta_i$ = 0 corresponds to a censored event. $t_i$ is the time to event or censoring. The overarching goal is to predict the survival risk $\mathbf r_i$ for each patient. The workflow is illustrated in Figure~\ref{Fig.slice}(a).



\subsection{Anchor slice selection}


First, we utilize slices from the three different planes, i.e the axial, coronal, and sagittal planes, to extract imaging features for each patient as shown in Figure \ref{Fig.slice}(b). We analyze the 3D volumes as 2D slices in multiple planes. For a given MRI, we calculate the number of tumor pixels and determine the slice indices $anchor_x$, $anchor_y$, $anchor_z$ corresponding to the largest tumor area in the sagittal, coronal, and axial planes, respectively. The corresponding slices are designated as anchor slices $s_{anchor_x}$, $s_{anchor_y}$, and $s_{anchor_z}$, respectively. Based on these anchor slices, we select neighbor slices as $[anchor_x-k_{x1},~anchor_x+k_{x2}],~[anchor_y-k_{y1},~anchor_y+k_{y2}],~[anchor_z-k_{z1},anchor_z+k_{z2}]$ near the anchor slices where $k_{q1},k_{q2}$ represent the number of left neighbors and right neighbors around $s_{anchor_q}$, q $\in$ (x,y,z). 
For each selected slice, we crop a bounding box around the tumor region then resize it into $224\times 224$. We now obtain a 2D slice list including the cropped slices in three planes for each MRI volume.


The number of slices $K$,  $K=k_{x1}+k_{x2}+k_{y1}+k_{y2}+k_{z1}+k_{z2}+3$, can influence the fraction of tumor selected. Higher the number of neighbor slices around the anchor slices, more is the volume of selected tumor region until the whole tumor is accounted for. The $ratio=\frac{\#\ tumor\ pixels\ in \ selected\ slices}{\#\ all\ tumor \ pixels }$ is shown in Figure \ref{Fig.slice}(c). With increase in $K$, the tumor pixels included across these slices also increases 
.We evaluate whether our choice of $K$ will influence the C-index.

\subsection{Finetuning Vision transformer on an MRI dataset}
We finetune a model on dataset $D_1$~\cite{Cheng2017} then utilize the trained model to extract features on dataset $D_2$. $D_1$ and $D_2$ are described in Section~\ref{d1andd2}.
Images in $D_1$ are fed into the ImageNet pretrained ViT for classification. For each image, the tumor region is cropped and resized to $224\times 224$ as the input to the pretrained model~\cite{rw2019timm}. The patch size is $16\times16$. The input channel is changed to 1 and the number of classes is changed to 3 according to the $D_1$ dataset. 500 epochs are training with Adam optimizer. We achieve a validation accuracy of 0.94 in classifying glioma, meningioma, and pituitary tumor.  Consequently, we obtain a ViT finetuned on MRI slices which is then leveraged as a feature extractor for survival prediction. Each slice from the slice list is fed into the extractor to get a slice-level representation. A list of slice-representations is obtained for each patient. 
\subsection{Deep anchor attention learning}
This section discusses the aggregation of slice-level representation into patient-level representation. Suppose one patient has $K$ slices, the corresponding slice-level representation list is $H=\{\mathbf h_1,\mathbf h_2,...,\mathbf h_K\}$, and the target is to generate the patient-level representation. 
As an aggregation strategy, we propose a Deep Anchor Attention Learning (DAAL) method to assign weights to each slice-level representation. 


Each slice-level representation $\mathbf h_s$ is transformed into query $\mathbf q_s\in \mathbb{R}^{D\times 1}$ and information $\mathbf v_s\in\mathbb{R}^{L\times 1}$ with:
\begin{equation}
    \mathbf q_s=\mathbf W_q\mathbf h_s,\quad \mathbf v_s=\mathbf W_v\mathbf h_s,\quad s=\{0,...,K-1\}\label{qv}
\end{equation}
where $\mathbf W_q$ and $\mathbf W_v$ are the weight matrices. The queries are to be matched between slices and the anchor slices. Information from slice-level representation are extracted using information vector $\mathbf v_s$.  The anchor slice representations are $\mathbf h_{anchor_x}$, $\mathbf h_{anchor_y}$, $\mathbf h_{anchor_z}$ and their corresponding query is $q_{anchor_p}(p=x,y,z).$The distance measurement $U$ between an arbitrary representation to the anchor slice representation is defined as:
\begin{equation}
\begin{aligned}
    U_p(\mathbf h_s,\mathbf h_{anchor_p})=\frac{exp(\left\langle\mathbf q_s,\mathbf q_{anchor_p}\right\rangle)}{\sum_{k=0}^{K-1}exp(\left\langle\mathbf q_k,\mathbf q_{anchor_p}\right\rangle)} \label{distance}
\end{aligned}
\end{equation}
where $\left\langle\cdot,\cdot\right\rangle$ represents the inner product of two vectors. It measures the similarity between the other queries (including anchor queries) and the anchor query. We set three distances $U_p$, $p=(x,y,z)$ according to the three anchor slices. The patient-level representation is the weighted element-wise sum of the information vectors $\mathbf v_i$ of all slice-level representations with the $U_p(\mathbf h_s,\mathbf h_{anchor_p})$ as weights:
\begin{equation}
\begin{aligned}
    b_p=\sum_s^{K-1}U_p(\mathbf h_s,\mathbf h_{anchor_p})\mathbf v_s\quad(p=x,y,z)
\end{aligned}
\end{equation}
We get three patient-level representations $b_x,b_y$, and $b_z$, which are then fed into fully connected layers to output the risk $r_{ix},r_{iy},r_{iz}$. Two experiments are performed to evaluate our methods. First,  we use only $r_{ix}$ as the patient risk score (DAAL-single). Second, we use $max(r_{ix},r_{iy},r_{iz})$ as the patient risk score.
We leverage the negative log partial likelihood function as the loss function in Equation \ref{eq:1} ~\cite{yao2020whole}.
\begin{equation}\label{eq:1}
    L(\mathbf r_i)=\sum_i\delta_i(-\mathbf r_i+log\sum_{j:t_j>=t_i}exp(\mathbf r_j))
\end{equation}

\section{Experiment Design and Results}

\subsection{Dataset Description}\label{d1andd2}

Dataset 1 ($D_1$) is a brain tumor dataset which includes 3064 T1-weighted contrast-enhanced images with three kinds of brain tumors, namely, glioma, meningioma, and pituitary tumor~\cite{Cheng2017}. It provides 2D slices in three planes along with associated tumor mask and tumor class. We leverage ($D_1$) for ViT fine tuning. 

To validate our performance, we utilized another dataset $D_2$ comprising N=326 MRIs and corresponding tumor masks from the BraTS2020 challenge~\cite{menze2014multimodal,bakas2017advancing,bakas2018identifying}. Available sequences include pre-contrast T1, post-contrast T1-weighted (T1Gd), T2-weighted, and T2 Fluid Attenuated Inversion Recovery (T2-FLAIR) volumes. The tumor annotations comprise the enhancing tumor, the peritumoral edema, and the necrotic and non-enhancing tumor core. We validate our proposed method on T1Gd scans. 49 patients are set aside as the test set according to the ratio of censored data. The remaining 277 patients are split into 5 folds for cross validation. Consequently we have five models in the 5-fold cross validation setting and test them on the held out test set.

\begin{table*}\small
\caption{C-index with Vision transformer feature extractor}
\begin{center}
\begin{tabular}{|l|c|c|c|c|c|c|c|c|c|c|}
\hline
Number of slices & 8 & 16 & 24 & 32 & 40 & 48 & 64 & 72 & 80 & 96 \\\hline
ViT+cox+mean&\textbf{0.6687}&0.6679&0.6719&0.6695&0.6641&0.6695&0.6747&0.6723&0.6719&0.6790\\\hline
ViT+cox+max&0.6623&0.6466&0.6344&0.6205&0.6191&0.6374&0.6729&0.6579&0.6754&\textbf{0.7029}\\\hline
ViT+pysurv+mean&0.6553&0.6649&0.6668&0.6667&0.6608&0.6730&0.6743&0.6863&0.6849&0.6844\\\hline
ViT+pysruv+max&0.6587&0.6556&0.6311&0.6623&0.6505&0.6642&0.6811&0.6842&0.6712&0.6679\\\hline
Attention MIL&0.6468&0.6567&0.6727&0.6774&0.6812&0.6866&0.6948&0.6928&0.6956&0.6940\\\hline
DAAL-single &0.6607&\textbf{0.6739}&\textbf{0.6766}&0.6838&\textbf{0.6894}&\textbf{0.6932}&\textbf{0.6986}&\textbf{0.6972}&\textbf{0.6996}&0.6992\\\hline
DAAL-multiple&0.6408&0.6525&0.6733&\textbf{0.6858}&0.6784&0.6860&0.6924&0.6950&0.6960&0.6916\\\hline
\end{tabular}
\end{center}
\label{daalvit}
\end{table*}

\begin{table*}\small
\caption{C-index with ResNet18 feature extractor}
\begin{center}
\begin{tabular}{|l|c|c|c|c|c|c|c|c|c|c|}
\hline
Number of slices & 8 & 16 & 24 & 32 & 40 & 48 & 64 & 72 & 80 & 96 \\\hline
ResNet18+cox+mean&0.6328&0.6374&0.6402&0.6396&0.6404&0.6430&0.6511&0.6515&0.6543&0.6591\\\hline
ResNet18+cox+max&0.6284&0.6396&0.6143&0.6243&0.6312&0.6352&0.6426&0.6456&0.6392&0.6436\\\hline
ResNet18+pysurv+mean&\textbf{0.6390}&\textbf{0.6516}&\textbf{0.6509}&0.6481&0.6491&0.6460&0.6503&0.6587&0.6578&0.6524\\\hline
ResNet18+pysurv+max&0.6277&0.6436&0.6332&0.6336&0.6404&0.6420&0.6431&0.6429&0.6388&0.6406\\\hline
Attention MIL&0.6354&0.6402&0.6394&0.6384&0.6424&0.6416&0.6488&0.6499&0.6529&0.6551\\\hline
DAAL-single &0.6348&0.6382&0.6474&\textbf{0.6500}&\textbf{0.6551}&\textbf{0.6547}&\textbf{0.6593}&\textbf{0.6607}&\textbf{0.6619}&\textbf{0.6609}\\\hline
DAAL-multiple&0.6274&0.6308&0.6382&0.6392&0.6460&0.6531&0.6482&0.6527&0.6541&0.6535\\\hline
\end{tabular}
\end{center}
\label{daalres}
\end{table*}
\subsection{Evaluation metrics}
Concordance index (C-index)~
is used to  evaluate the prognostic models. 
A c-index $>$ 0.65-0.7 is generally considered optimal in survival analysis.
The hazard ratio (HR) is the ratio of the hazard rates corresponding to the conditions described by two levels of an explanatory variable.
We use HR to compare the probability of death events in two groups (low risk and high risk groups). For each fold, we obtain a risk score for the patients in the test set. The two risk groups are identified by the median risk score from the training set. We then use majority voting for final risk group assignment.
\vspace{-4mm}
\subsection{Comparisons}
\paragraph{Radiomics.} Many prior image-based survival analysis methods have focused on  radiomics. We extract 336 textural radiomic measures from the intratumoral regions using the provided tumor segmentation~\cite{van2017computational} of the T1-Gd MRIs follwed by a Principal Component Analysis~\cite{wold1987principal}. We use the Cox model with 10 principal components as input. The C-index is 0.5796 ± 0.0309.

\paragraph{ResNet18.} In order to evaluate the difference between features from a ViT and other CNN models, we leverage a pre-trained ResNet18~\cite{he2016deep} finetuned on a brain tumor dataset to keep the setup consistent with our ViT experiments. 512 features are extracted for each slice followed by our proposed DAAL method to evaluate the survival prediction results.


\paragraph{Cox model.}
The Cox proportional hazards model is the most popular model in survival analysis~\cite{choi2020multi}. After computation of the ViT features $\mathbf H=\{\mathbf h_1,\mathbf h_2,...,\mathbf h_K\}$, we calculate the mean $\mathbf H_{mean}$ and maximum $\mathbf H_{max}$ across the slice-level representations. $\mathbf H_{mean}$ and $\mathbf H_{max}$ also work as the patient-level representations and are fed into the Cox model. We also feed the mean and maximum values of ResNet18 representation to a Cox model for comparison.

\paragraph{DeepSurv.}
DeepSurv~\cite{katzman2018deepsurv} is another commonly used survival prediction model. We use the $\mathbf H_{mean}$ and  $\mathbf H_{max}$ to be the patient level-representation and feed them into the DeepSurv model from the pysurv package~\cite{pysurvival_cite}.

\paragraph{DeepAttnMISL model.} DeepAttnMISL~\cite{yao2020whole} is a deep attention multiple instance learning method first proposed for whole slide tissue images. We use the attention MIL~\cite{ilse2018attention} method from DeepAttnMISL to assign the weights to ViT slice-level representations and ResNet18 slice-level representations to compare the attention MIL with DAAL.
\subsection{Survival Analysis Results}

 Table \ref{daalvit} shows the results of our methods (DAAL-single, DAAL-multiple) and attention MIL using vision transformer feature extractor. We also consider the aggregation methods with the mean and max values of the slice-level representation with Cox model. We see that in most cases, our methods show higher average C-index values compared with attention MIL and Cox models. More importantly, the proposed methods show \textit{higher C-index values even with small number of slices} as compared to attention MIL and Cox models, implying that we can predict survival risk reliably using relatively less contextual information.
\begin{figure}
	\centering
   \subfigure[HR with ViT]{
		\begin{minipage}[b]{0.19\textwidth}
			\includegraphics[width=1\textwidth]{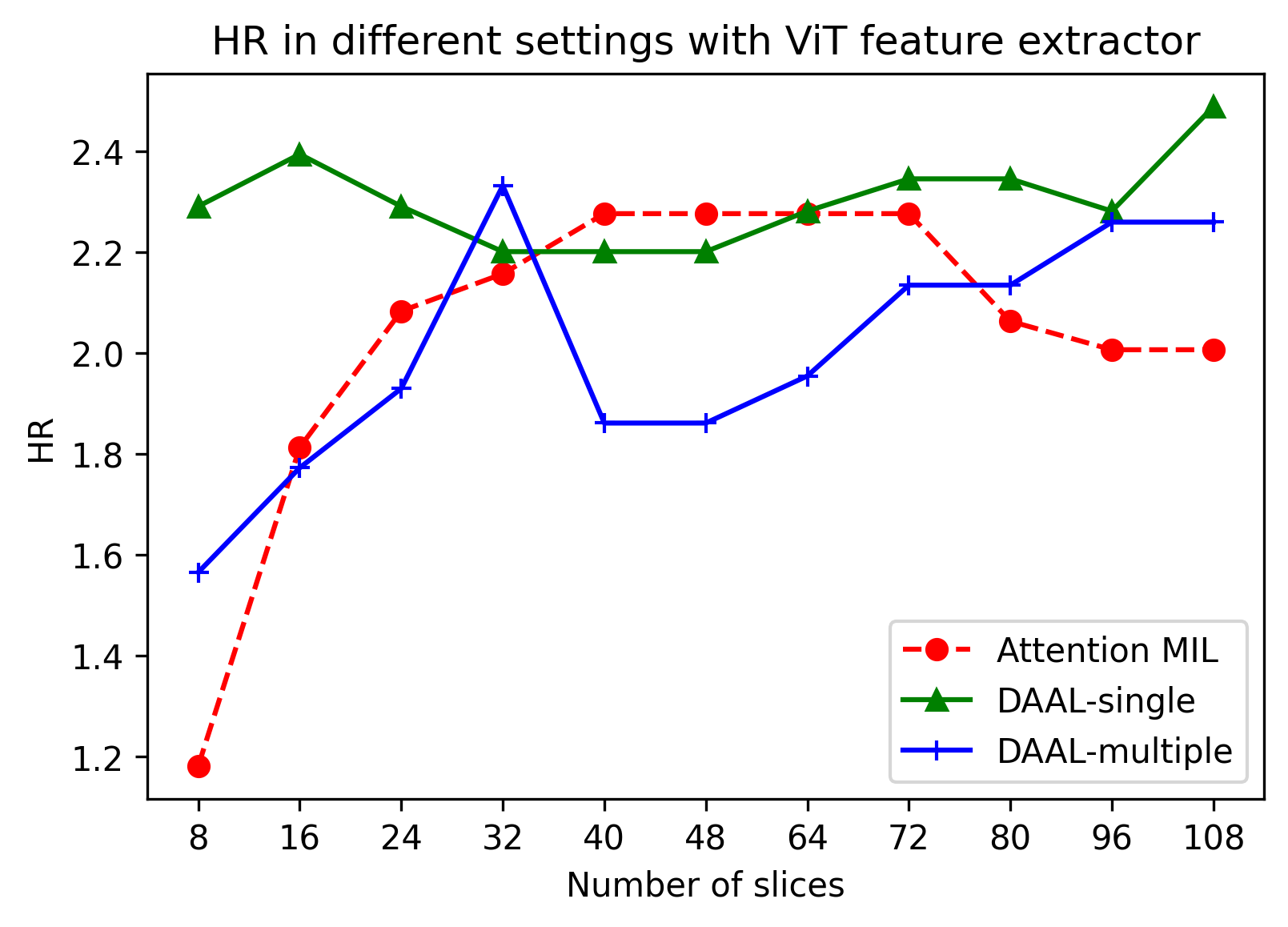}
		\end{minipage}
		\label{hrvitmil}
	}
    \subfigure[HR with ResNet18]{
    		\begin{minipage}[b]{0.19\textwidth}
   		 	\includegraphics[width=1\textwidth]{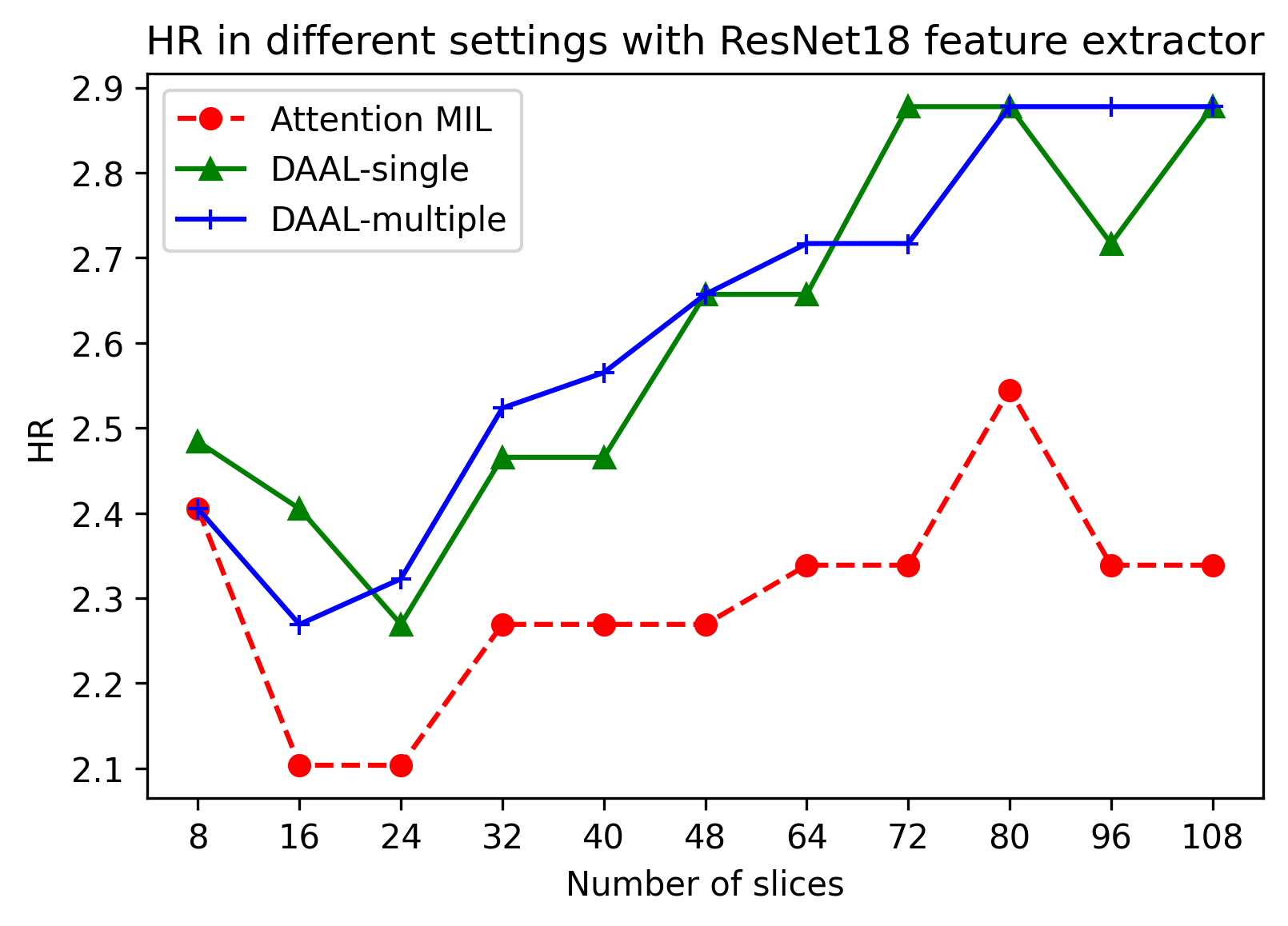}
    		\end{minipage}
		\label{resc}
    	}
	\caption{Hazard ratio distribution in different settings.}
	\label{hrresmil}
\label{cindexcomp}
\end{figure}

Besides ViT, we also evaluate these methods with a ResNet18 feature extractor. Table \ref{daalres}  show the c-indices with ResNet18 features. From slice number $K=32$, DAAL-single shows higher c-indices compared to all the other methods with ResNet18 features. We compare the deep learning methods (Attention MIL, DAAL-single, and DAAL-multiple) in Figure~\ref{cindexcomp}. As may be observed, using ViT we get higher c-indices as compared to ResNet18 at identical slice number settings. The results suggest intra-slice spatial diversity is inherently prognostic and can be captured by self attention.

\section{Discussion and Conclusion}

In this work, we proposed a deep anchor attention learning method with a vision transformer and demonstrated its efficacy in predicting overall survival in brain cancer patients using treatment na\"ive MRI. The intra-slice spatial diversity is captured by the vision transformer.
The DAAL method is used as an aggregation strategy; this highlights the important slices and estimates the inter-slice spatial diversity. The approach implicitly mimics the experts’ diagnostic process, where radiologists usually observe slices with more well defined and large tumor components first and then focus their attention on the other slices (may be considered secondary, yet valuable to the ultimate diagnosis) to evaluate disease severity.
 Experiments show that DAAL leads to an improvement in survival prediction over state-of-the-art methods. Our methods achieve higher c-indices with lower contextual information. 
In the future, we may consider anchor attention mechanisms in 3D models and 
complementing T1-Gd with other sequences (T2, FLAIR) when available.





\section{Compliance with ethical standards}
\label{sec:ethics}

This  research  study  was  conducted  retrospectively  using open access human subject data (BraTS 2020). Additional approval was not required as confirmed by~\cite{menze2014multimodal,bakas2017advancing,bakas2018identifying}.

\section{Acknowledgments}
\label{sec:acknowledgments}
No funding was received for conducting this study. The authors have no financial or non-financial interests to disclose.

\bibliographystyle{IEEEbib}
\bibliography{strings,refs,egbib}

\end{document}